\documentclass[journal=ancac3,manuscript=article]{achemso}
\setkeys{acs}{articletitle=true,etalmode=truncate,maxauthors=100}

\usepackage[version=3]{mhchem} 
\usepackage[T1]{fontenc}       

\usepackage{color}
\usepackage[table]{xcolor}
\usepackage{float}
\usepackage[normalem]{ulem}
\usepackage{hyperref}
\usepackage{makecell}
\usepackage{array}

\newcommand{\White}[1]{\textcolor{white}{#1}}

\author{Anike Purbawati}
\affiliation{Univ. Grenoble Alpes, CNRS, Grenoble INP, Institut NEEL, 38000 Grenoble, France}
\altaffiliation{Contributed equally to this work}
\author{Suman Sarkar}
\affiliation{Univ. Grenoble Alpes, CNRS, Grenoble INP, Institut NEEL, 38000 Grenoble, France}
\altaffiliation{Contributed equally to this work}
\author{S\'{e}bastien Pairis}
\affiliation{Univ. Grenoble Alpes, CNRS, Grenoble INP, Institut NEEL, 38000 Grenoble, France}
\author{Marek Kostka}
\affiliation{Univ. Grenoble Alpes, CNRS, Grenoble INP, Institut NEEL, 38000 Grenoble, France}
\alsoaffiliation{Institute of Physical Engineering, Brno University of Technology, Brno 616 69, Czech Republic}
\author{Abdellali Hadj-Azzem}
\affiliation{Univ. Grenoble Alpes, CNRS, Grenoble INP, Institut NEEL, 38000 Grenoble, France}
\author{Didier Dufeu}
\affiliation{Univ. Grenoble Alpes, CNRS, Grenoble INP, Institut NEEL, 38000 Grenoble, France}
\author{Priyank Singh}
\affiliation{Univ. Grenoble Alpes, CNRS, Grenoble INP, Institut NEEL, 38000 Grenoble, France}
\author{Daniel Bourgault}
\affiliation{Univ. Grenoble Alpes, CNRS, Grenoble INP, Institut NEEL, 38000 Grenoble, France}
\author{Manuel Nu\~{n}ez-Regueiro}
\affiliation{Univ. Grenoble Alpes, CNRS, Grenoble INP, Institut NEEL, 38000 Grenoble, France}
\author{Jan Vogel}
\affiliation{Univ. Grenoble Alpes, CNRS, Grenoble INP, Institut NEEL, 38000 Grenoble, France}
\author{Julien Renard}
\affiliation{Univ. Grenoble Alpes, CNRS, Grenoble INP, Institut NEEL, 38000 Grenoble, France}
\author{La\"{e}titia Marty}
\affiliation{Univ. Grenoble Alpes, CNRS, Grenoble INP, Institut NEEL, 38000 Grenoble, France}

\author{Florentin Fabre}
\affiliation{Laboratoire Charles Coulomb, Universit\'{e} de Montpellier and CNRS, 34095 Montpellier, France}
\author{Aurore Finco}
\affiliation{Laboratoire Charles Coulomb, Universit\'{e} de Montpellier and CNRS, 34095 Montpellier, France}
\author{Vincent Jacques}
\affiliation{Laboratoire Charles Coulomb, Universit\'{e} de Montpellier and CNRS, 34095 Montpellier, France}

\author{Lei Ren}
\affiliation{Universit\'{e} de Toulouse, INSA-CNRS-UPS, LPCNO, 135 Av. Rangueil, 31077 Toulouse, France}
\author{Vivekanand Tiwari}
\affiliation{Universit\'{e} de Toulouse, INSA-CNRS-UPS, LPCNO, 135 Av. Rangueil, 31077 Toulouse, France}
\author{Cedric Robert}
\affiliation{Universit\'{e} de Toulouse, INSA-CNRS-UPS, LPCNO, 135 Av. Rangueil, 31077 Toulouse, France}
\author{Xavier Marie}
\affiliation{Universit\'{e} de Toulouse, INSA-CNRS-UPS, LPCNO, 135 Av. Rangueil, 31077 Toulouse, France}

\author{Nedjma Bendiab}
\affiliation{Universit\'{e} Grenoble Alpes, CNRS, Grenoble INP, Institut NEEL, 38000 Grenoble, France}
\author{Nicolas Rougemaille}
\affiliation{Universit\'{e} Grenoble Alpes, CNRS, Grenoble INP, Institut NEEL, 38000 Grenoble, France}
\author{Johann Coraux}
\affiliation{Universit\'{e} Grenoble Alpes, CNRS, Grenoble INP, Institut NEEL, 38000 Grenoble, France}
\email{johann.coraux@neel.cnrs.fr}

\title[Stability of CrTe$_2$]
  {Stability of the In-Plane Room Temperature van der Waals Ferromagnet Chromium Ditelluride and Its Conversion to Chromium-Interleaved CrTe$_2$ Compounds}


\begin{document}



\bigskip
\bigskip

\begin{abstract}

Van der Waals magnetic materials are building blocks for novel kinds of spintronic devices and playgrounds for exploring collective magnetic phenomena down to the two-dimensional limit. Chromium-tellurium compounds are relevant in this perspective. In particular, the 1$T$ phase of CrTe$_2$ has been argued to have a Curie temperature above 300~K, a rare and desirable property in the class of lamellar materials, making it a candidate for practical applications. However, recent literature reveals a strong variability in the reported properties, including magnetic ones. Using electron microscopy, diffraction and spectroscopy techniques, together with local and macroscopic magnetometry approaches, our work sheds new light on the structural, chemical and magnetic properties of bulk 1$T$-CrTe$_2$ exfoliated in the form of flakes having a thickness ranging from few to several tens of nanometers. We unambiguously establish that 1$T$-CrTe$_2$ flakes are ferromagnetic above room temperature, have an in-plane easy axis of magnetization, low coercivity, and we confirm that their Raman spectroscopy signatures are two modes, $E_{2\text{g}}$ (103.5~cm$^{-1}$) and $A_{1\text{g}}$ (136.5~cm$^{-1}$). We also prove that thermal annealing causes a phase transformation to monoclinic Cr$_5$Te$_8$ and, to a lesser extent, to trigonal Cr$_5$Te$_8$. In sharp contrast with 1$T$-CrTe$_2$, none of these compounds have a Curie temperature above room temperature, and they both have perpendicular magnetic anisotropy. Our findings reconcile the apparently conflicting reports in the literature and open opportunities for phase-engineered magnetic properties.

\end{abstract}



\newpage

\section*{Introduction}

Crystalline magnetic materials with layered structure, preferably held together by interlayer van der Waals forces, have been arousing considerable interest since it was realised they can be thinned down to single layers and still host (sometimes unusual) ordered magnetic phases \cite{Lee,Wang_e,Huang_b,Gong,Gibertini,Wang_rev}. These materials can be used as stamps to assemble new spintronic architectures \cite{Gibertini,Wang_rev}, with which concepts originally introduced for conventional magnets (\textit{i.e.} not van der Waals) can be revisited. This includes chiral magnetization textures \cite{Han,Ding,Park,Sun_c}, magnonic excitations \cite{Cenker}, and interface effects controlled with electric fields \cite{Xing,Deng,Jiang,Jiang_b,Wang} or using the proximity to other functional materials \cite{Ciorciaro,Zhang}.

Practical spintronic applications generally require materials with Curie temperature, $T_\mathrm{Curie}$, above 300~K, which is not the case for most van der Waals ferromagnets. One strategy is to increase $T_\mathrm{Curie}$ of known materials by modifying their structure/composition via chemical substitution \cite{Tian} or ion implantation \cite{Yang}, or by influencing them by proximity effect \cite{Wang_c}. Complementary to this strategy, the search for alternative materials is another, potentially guided by predictions of \textit{ad hoc} chemical compositions and structures. This implies the exploration of sometimes complex alloy phase diagrams using various synthesis methods and multi-technique characterisations. In this context Fe-Ge-Te and Cr-Te compounds have attracted most attention, and we address the latter category in the present work.

Structure and composition control in the Cr-Te binary system is a known delicate problem. Indeed a variety of compounds with different stoichiometry can be formed, including CrTe \cite{Wang_b} ($T_\mathrm{Curie}\sim$200~K), Cr$_2$Te$_3$ \cite{Hashimoto,Bian,Zhong} ($T_\mathrm{Curie}\sim$170~K), Cr$_{1+1/3}$Te$_2$ \cite{Lasek} ($T_\mathrm{Curie}\sim$160-190~K), Cr$_3$Te$_4$ \cite{Yamaguchi,Chua} ($T_\mathrm{Curie}\sim$240-320~K), Cr$_3$Te$_5$ \cite{Huang}, Cr$_4$Te$_5$ \cite{Zhang_c} (both with a $T_\mathrm{Curie}\sim$320~K), and Cr$_5$Te$_8$ \cite{Wang_d,Chen,Tang} ($T_\mathrm{Curie}\sim$100-220~K). None of these binary alloys are lamellar \textit{per se}, but are thought to consist of CrTe$_2$ layers where Cr atoms have octahedral local environment with covalent Cr bridges between these layers. The magnetic properties might be strongly affected by the density of such bridges, and so should exfoliation, which is expectedly easier in the case of van der Waals interlayer interactions.

\begin{figure}[!hbt]
 \begin{center}
 \includegraphics[width=8cm]{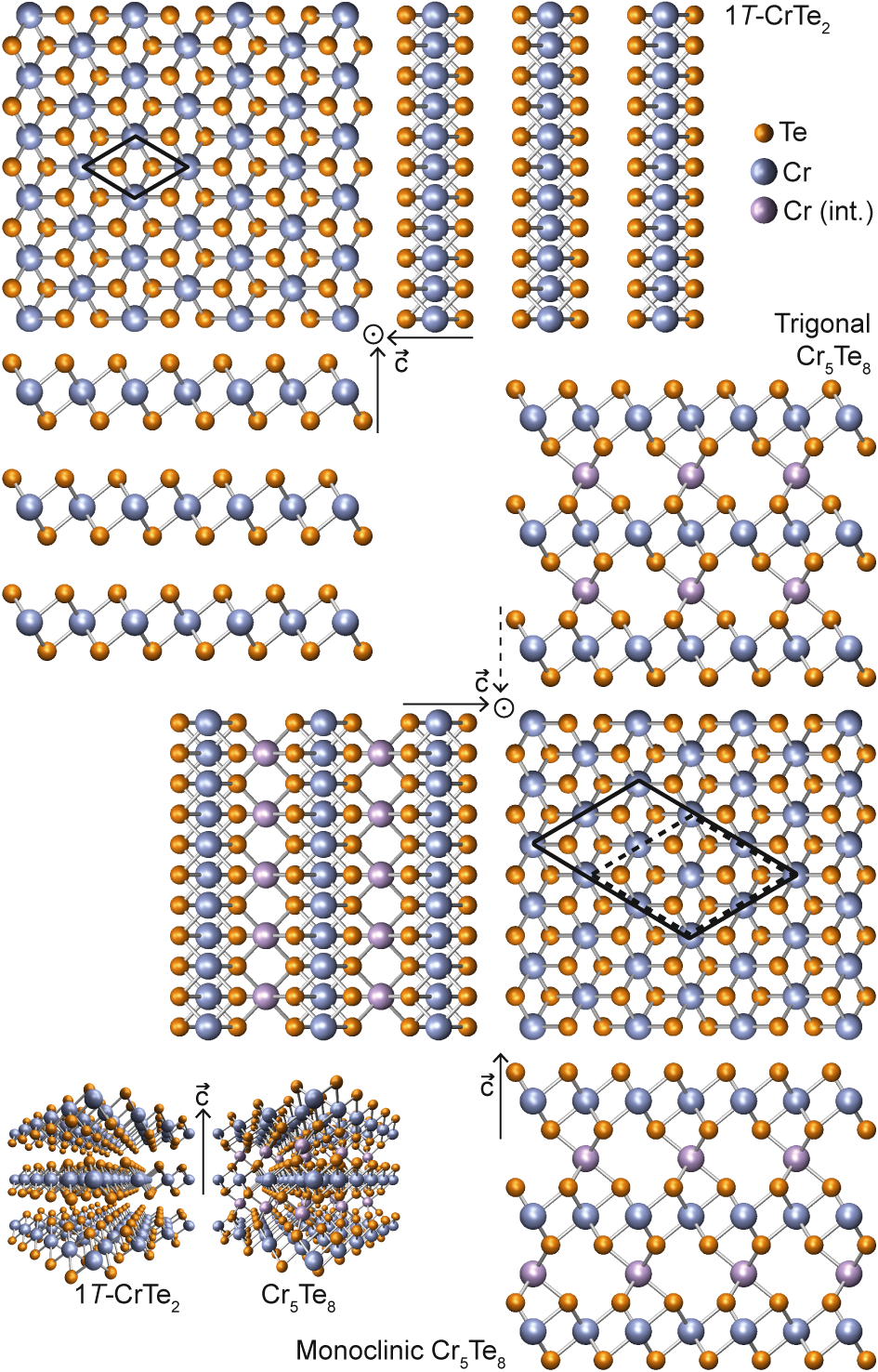}
 \caption{\label{fig0} \textbf{Atomic structure of Cr-Te compounds.} Ball-and-stick models of 1$T$-CrTe$_2$ (top) and Cr$_5$Te$_8$ (bottom), on top views and side views, along two perpendicular crystal axes. Two possible structures are shown for Cr$_5$Te$_8$, one trigonal and the other monoclinic. A three-dimensional view of the lattice of 1$T$-CrTe$_2$ and Cr$_5$Te$_8$ is also shown at the bottom left. The unit cells are sketched on the top views with solid (1$T$-CrTe$_2$, monoclinic Cr$_5$Te$_8$) or dotted (trigonal Cr$_5$Te$_8$) lines.}
 \end{center}
\end{figure}

This weak interaction should prevail only for the 1:2 Cr:Te composition, in CrTe$_2$ crystallised in the $1T$ polytype (\autoref{fig0}). Several attempts have been made recently to prepare this compound in the form of (ultra)thin films. High temperature synthesis of the potassium-intercalated bulk material, followed by de-intercalation and mechanical exfoliation \cite{Freitas,Purbawati,Sun}, large-area bottom-up growth by molecular beam epitaxy on GaAs \cite{Sun_b} and graphene \cite{Zhang_b}, and chemical vapour deposition on SiO$_2$\cite{Meng}, have been reported. Variations of the magnetic properties as function of the thickness of the material have been observed: decrease of $T_\mathrm{Curie}$ of the order of 10~K (around few nanometers to few tens of nanometers \cite{Purbawati,Sun,Fabre,Meng,Sun_b}) and even 100~K (for the single-layer \cite{Zhang_b}), reorientation of the magnetization easy axis (in-plane to out-of-plane below 10~nm \cite{Meng}), reduction of the coercivity \cite{Sun_b,Purbawati}, transition to an antiferromagnetic order for a single layer \cite{Xian}. This could signal a transition from a bulk to a two-dimensional character; unfortunately this corpus of recent results is not consistent.

Beyond this, puzzling discrepancies and even contradictions have appeared, although they are sometimes difficult to spot because different sizes, shapes and thicknesses are considered in the reports. In the case of the thick material produced by each method however, the variability is better identified. The state of affairs is indeed confusing: $T_\mathrm{Curie}$ slightly above 300~K \cite{Freitas,Purbawati,Sun,Fabre,Zhang_b} or $\sim$200~K \cite{Meng,Sun_b}, in-plane \cite{Freitas,Purbawati,Sun,Fabre,Meng} and out-of-plane \cite{Sun_b,Zhang_b} easy magnetization axis, and (in-plane) coercivity of several 1~mT \cite{Purbawati,Sun} and several 100~mT \cite{Meng}, have been reported. Why should the preparation method influence these properties?

Here, we address this question under the prism of the material stability, which is a major issue for almost all van der Waals magnetic materials so far. We consider the impact of sample annealing on its structure, composition and magnetic properties, while the effect of air exposure is marginal for timescales of at least hours. We first thoroughly and quantitatively characterize pristine $1T$-CrTe$_2$ synthesised at high temperature in the bulk form and subsequently exfoliated, with diffraction and element-sensitive micro-analysis tools. Next, we relate these analyses to a detailed Raman spectroscopy characterization, in an attempt to clarify the rather contradicting results found in the literature. We further quantitatively show that thermal annealing at 500~K and 800~K induces a change of the structural and chemical composition, to Cr$_5$Te$_8$. We correlate this information to drastic changes of the magnetic properties. This allows us to rationalise the various behaviours reported in the existing literature: the variability in the observed magnetic properties can be ascribed to varying Cr:Te compositions. Our work shows that while pristine $1T$-CrTe$_2$ is a room temperature ferromagnet with in-plane easy axis in the form of macroscopic grains and thin flakes, it transforms into other Cr-Te compounds upon thermal annealing, concomitantly loosing room temperature ferromagnetism and exhibiting an out-of-plane magnetic anisotropy.

\section*{Results and Discussion}

\subsection*{Methodology and strategy}

We study materials synthesized \textit{via} prolonged high temperature treatment of a 1:1:2 mixture of K, Cr, Te, and subsequent slow cool down. As described in \textit{Materials and Methods} and Refs.~\citenum{Freitas,Sun}, special attention was paid to prevent oxydation by using an Ar-filled glove box, both to fill the quartz cell used for growth and to de-intercalate the grown material from K atoms. In the following we discuss results obtained (i) on bulk samples, \textit{i.e.} macroscopic millimeter-sized grains of thickness of the order of hundreds of micrometers, and (ii) on flakes having an extension of the order of 10~$\mu$m and thicknesses in the range of few tens of nanometers, which we deposited on Pt thin films (see Table~\ref{tab:samples} and \textit{Materials and Methods}). Macroscopic samples were used to explore the magnetic properties as a function of temperature, down to 10~K with a superconducting quantum interference device (SQUID). Micro-flakes were probed at room temperature, in a local fashion under the objectives of optical (Raman spectroscopy and focussed Kerr magnetometry) and electron (energy-dispersive X-ray spectroscopy -- EDS -- and electron backscatter diffraction -- EBSD --) microscopes. This allowed us to correlate complementary observations unambiguously.

First, our goal was to resolve the crystal symmetry and composition, as a function of an annealing temperature (500~K, 650~K and 800~K, annealing time: 30~min at each temperature). Second, we determined the magnetic properties of the sample in its pristine form (un-annealed) and after successive annealings. We then compare our observations to those found in the literature.

\begin{table*}[!hbt]
\begin{tabular}{p{1.3cm}|p{2.1cm}|p{1.9cm}|p{1.3cm}p{1.3cm}p{1.4cm}p{0.9cm}|p{1.7cm}}
\rowcolor{gray}\White{Sample} & \White{Dimensions} & \White{Annealing} & \White{Raman} & \White{EBSD} & \White{SQUID} & \White{Kerr} & \White{Figure \#} \\
\rowcolor{gray}\White{type} &  &  &  & \White{+EDS} &  &  &  \\
\hline
Macro. & $\sim$100~$\mu$m & none & yes & no & yes & no & \ref{fig3},\ref{fig4} \\
 & $\times$1~mm$^2$ & 500~K & yes & no & yes & no & \\
 & & 650~K & yes & no & yes & no & \\
 & & 800~K & yes & no & yes & no & \\

\rowcolor{lightgray}Micro- & 85~nm$^2$ & none & yes & yes & no & yes & \ref{fig1},\ref{fig2},\ref{fig3},\ref{fig4}\\
\rowcolor{lightgray}flake & $\times\sim$35~$\mu$m$^2$ & 500~K & yes & no & no & yes & \\
\rowcolor{lightgray} & & 800~K & yes & yes & no & yes & \\

Micro- & 53~nm$^2$ & none & yes & yes & no & yes & \ref{fig1},\ref{fig3},\ref{fig4}\\
flake & $\times\sim$80~$\mu$m$^2$ & 500~K & yes & yes & no & yes & \\
 & & 800~K & yes & yes & no & yes & \\

\rowcolor{lightgray}Micro- & 150~nm$^2$ & 500~K & no & yes & no & yes & S3,S4\\
\rowcolor{lightgray}flake & $\times\sim$12~$\mu$m$^2$ & & & & & & \\

\end{tabular}
\caption{Samples studied in this work (micro-flakes or macroscopic grains, in short `macro'), initially in the 1$T$-CrTe$_2$ phase, their thermal annealing temperatures, and the experimental techniques (Raman spectroscopy, EBSD+EDS, SQUID magnetometry, focussed Kerr magnetimetry) used to investigate them, whose analysis is presented in this article.}
\label{tab:samples}
\end{table*}

\subsection*{Crystallography and chemical analysis of pristine 1$T$-CrTe$_2$ micro-flakes} 
\begin{figure*}[!hbt]
 \begin{center}
 \includegraphics[width=16cm]{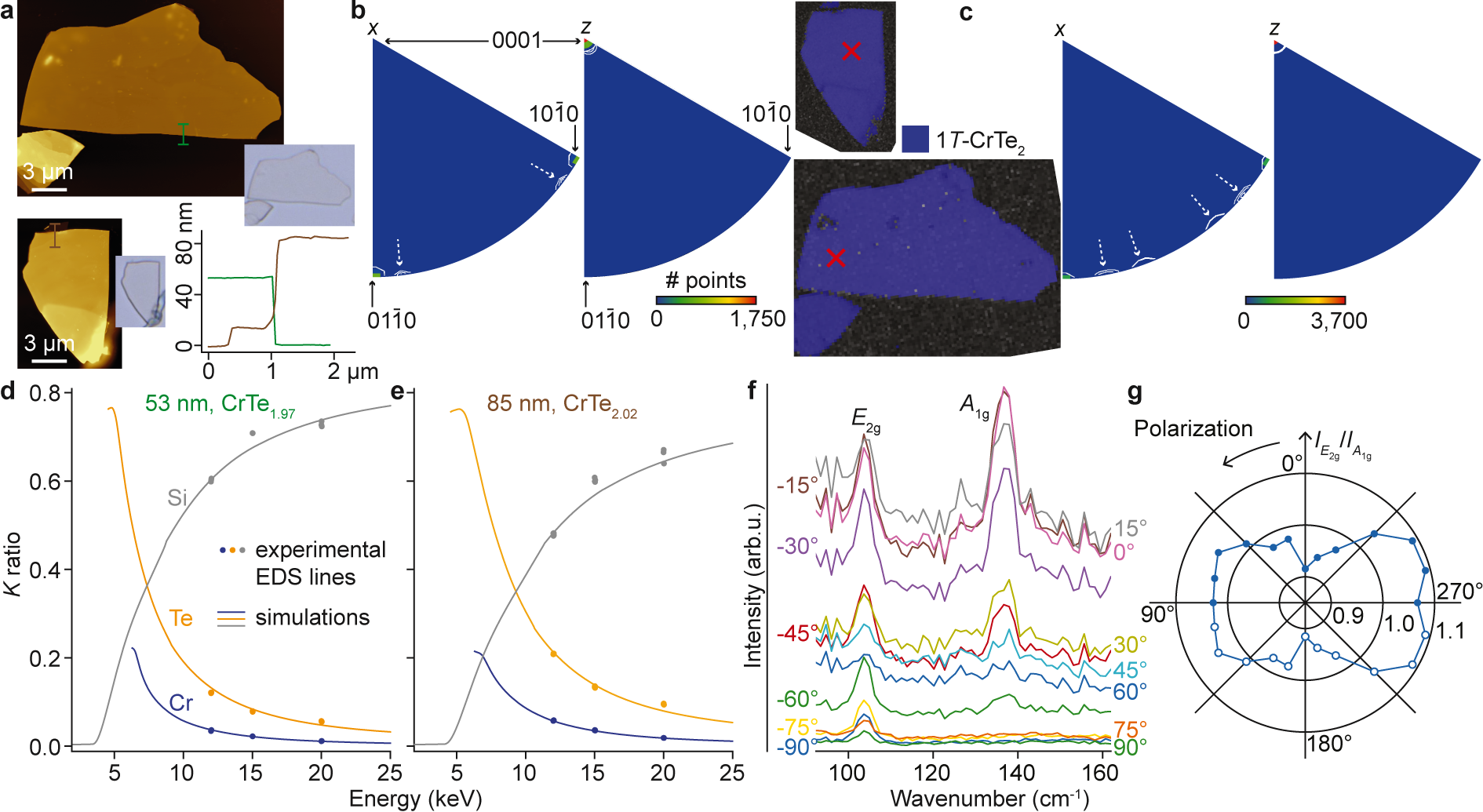}
 \caption{\label{fig1} \textbf{Composition, symmetry and vibrational properties of pristine 1$T$-CrTe$_2$ micro-flakes.} (a) AFM and optical microscopy of two exfoliated flakes placed onto a Pt surface, together with AFM height profiles. (b,c) The inverse pole figures (60$^\circ$ sectors) from EBSD are shown for these flakes along two axes (in-plane $x$, out-of-plane $z$). The corresponding structure assignment with SEM-EBSD (overlaid on the point-by-point band contrast of the Kikuchi line patterns across the SEM image; 1$T$-CrTe$_2$ phase appearing in blue; points with no successful assignment have no color) is shown on the areas mapped in (b). Projected reflections accumulating most fractions (in number of points) of the SEM images are found at the corners of the 60$^\circ$ sectors; dotted arrows highlight minority contributions from smaller flakes. (d,e) EDS analysis of the $K$-ratios for the Cr, Te, Si emission lines (coloured dots) at specific locations marked with crosses on (b,c), at different electron energies, together with simulated data. (f) Polarized Raman spectra of a pristine flake sealed under Ar atmosphere inside a quartz cell. (g) Polar plot of the ratio of intensity ratio of the $E_{2\text{g}}$ and $A_{1\text{g}}$ modes (103.5 and 136.5~cm$^{-1}$ respectively). The data have been symmetrized to produce the [90$^\circ$-270$^\circ$] sector.}
 \end{center}
\end{figure*}

Freshly exfoliated micro-flakes (see two examples in \autoref{fig1}a, 53~nm- and 85~nm-thick flakes as measured with atomic force microscopy -- AFM) were introduced in a scanning electron microscope (SEM). Inverse pole figures deduced from electron backscatter diffraction (EBSD, see \textit{Materials and Methods} and an example pattern in Supporting Information Figure~S1) reveal a perfect orientation of the crystal's $[0001]$ direction perpendicular to the flakes' surface (\textit{i.e.}, to the layers) and overall patterns (\autoref{fig1}b, for two axes) indexed assuming a P$\bar{3}m$1 (\#164) space group. In these patterns, two main contributions are observed in the $\lbrace12\bar{3}0\rbrace$ directions ($x$ axis) and a single one in the $[0001]$ direction ($z$ axis) (see Supporting Information, note in Section~S1). The patterns correspond to the expected $1T$ polytype of CrTe$_2$. The nature of the crystallographic phase can be determined point-by-point across the flakes' surface within the SEM image: our pristine flakes are all uniformly indexed with the same structure (uniform color on the flakes' mapping presented in \autoref{fig1}b,c, see Supporting Information Table~S1 for details about alternative candidate phases considered). Note that the crystal lattice of other Cr-Te compounds shares several symmetry elements and might not seem straightforwardly different in diffraction experiments. Careful analysis of the local lattice constants is necessary to draw conclusions, which we did here. The issue of structure resolution of Cr-Te compounds is all the more important as direct observation of the atomic structure by plane-view high-resolution transmission electron microscopy (see \textit{e.g.} Refs.~\citenum{Sun,Chen}) can in principle not differentiate the various compounds since Cr interlayer bridges, when they are present, are aligned with the Cr atoms within the CrTe$_2$ layers.


As already stressed in recent reports \cite{Huang,Chen}, extensive composition analysis is crucial. For that purpose EDS was used, via datasets acquired at random locations on the flakes (five to 10 per flake). In our EDS analysis, given their relatively low thickness the micro-flakes are semi-transparent to the electron beam, which therefore is scattered by the substrate. To evaluate this effect and take into account the finite thickness of the flakes we acquired data at three different electron energies, extracted the so called $K$-ratios of intensities from individual EDS spectra such as the one shown in Figure~S1, relative to standard values (see \textit{Materials and Methods}) for different X-ray emission lines (\autoref{fig1}d,e). Next, the $K$-ratios were simulated for varying compositions at the different energies, and best fit to the experimental $K$-ratios was sought for considering all energy values simultaneously (see \autoref{fig1}d,e for examples of the outcome of local analysis performed at the spots marked with crosses in \autoref{fig1}b,c). The most prominent emission lines are those of Cr, Te, and Si. We also sought for a potassium signature, indicative of partial de-intercalation of the bulk compound. However, sensitivity is not optimal given the close proximity of the Te emission line. Yet, our analysis indicates that the K content is at most 1\% (mass percent), possibly even less. The analysis of ten flakes yields a characteristic Te/Cr ratio of atomic percentage composition of 2.00$\pm$0.03, in remarkable agreement with the independent assignment made based on inverse pole figures. This ratio represents a so far unreported (to our best knowledge) clear evidence of a 1:2 Cr:Te composition. Note that our analysis revealed no substantial presence of oxygen.

\subsection*{Raman spectroscopy of pristine 1$T$-CrTe$_2$ micro-flakes} Having determined the structure and composition of the material, we now provide a Raman characterisation that will be used to track possible changes in the nature of the material. As shown in our previous work \cite{Purbawati}, two peaks are observed in the spectra, centered at 103.5~cm$^{-1}$ and 136.5~cm$^{-1}$, whatever the flake thickness in the range we explored (10-150~nm). The spectra are identical for flakes exfoliated within a glove box and then sealed within a quartz cell filled with Ar (for Raman spectroscopy measurements) and samples probed in atmospheric conditions even after prolonged exposure to air.

Based on a coarse analysis of the polarization of the two peaks, we previously ascribed the two peaks to the $E_{2\text{g}}$ (103.5~cm$^{-1}$) and $A_{1\text{g}}$ (136.5~cm$^{-1}$) modes \cite{Purbawati}. Here we provide a more detailled analysis, by continuously varying the polarization angle (\autoref{fig1}f). The intensity ratio between the $E_{2\text{g}}$ and $A_{1\text{g}}$ modes is found to vary with a 180$^\circ$ periodicity (\autoref{fig1}g). The highest-wavenumber peak having maximum intensity in the $Z(XX)\bar{Z}$ polarization configuration and vanishing completely in the $Z(XY)\bar{Z}$ configuration. The lowest-wavenumber peak also looses intensity in the latter configuration, but does not vanish. This behaviour is reminiscent of that observed in other transition metal dichalcogenide (\textit{e.g.} WS$_2$ \cite{Sekine}), and allows us to discriminate the two modes, here in-plane $E_{2\text{g}}$ and out-of-plane $A_{1\text{g}}$).

We note that Meng \textit{et al.} observed two peaks, centered at 124~cm$^{-1}$ and 144~cm$^{-1}$, which they assigned to $E_{2\text{g}}$ and $A_{1\text{g}}$ modes in CrTe$_2$ \cite{Meng}. This is inconsistent with our measurements and phase assignment. Also interesting is the observation of two peaks centered around 124~cm$^{-1}$ and 144~cm$^{-1}$ in Raman spectra measured for CrTe \cite{Wang_b}, Cr$_5$Te$_8$ \cite{Chen,Fu}, and Cr$_2$Te$_3$ \cite{Zhong}. We will shortly come back to this point.

\subsection*{Thermally-induced transformation of 1$T$-CrTe$_2$} 
\begin{figure}[!hbt]
 \begin{center}
 \vspace{-10pt}
 \includegraphics[width=8cm]{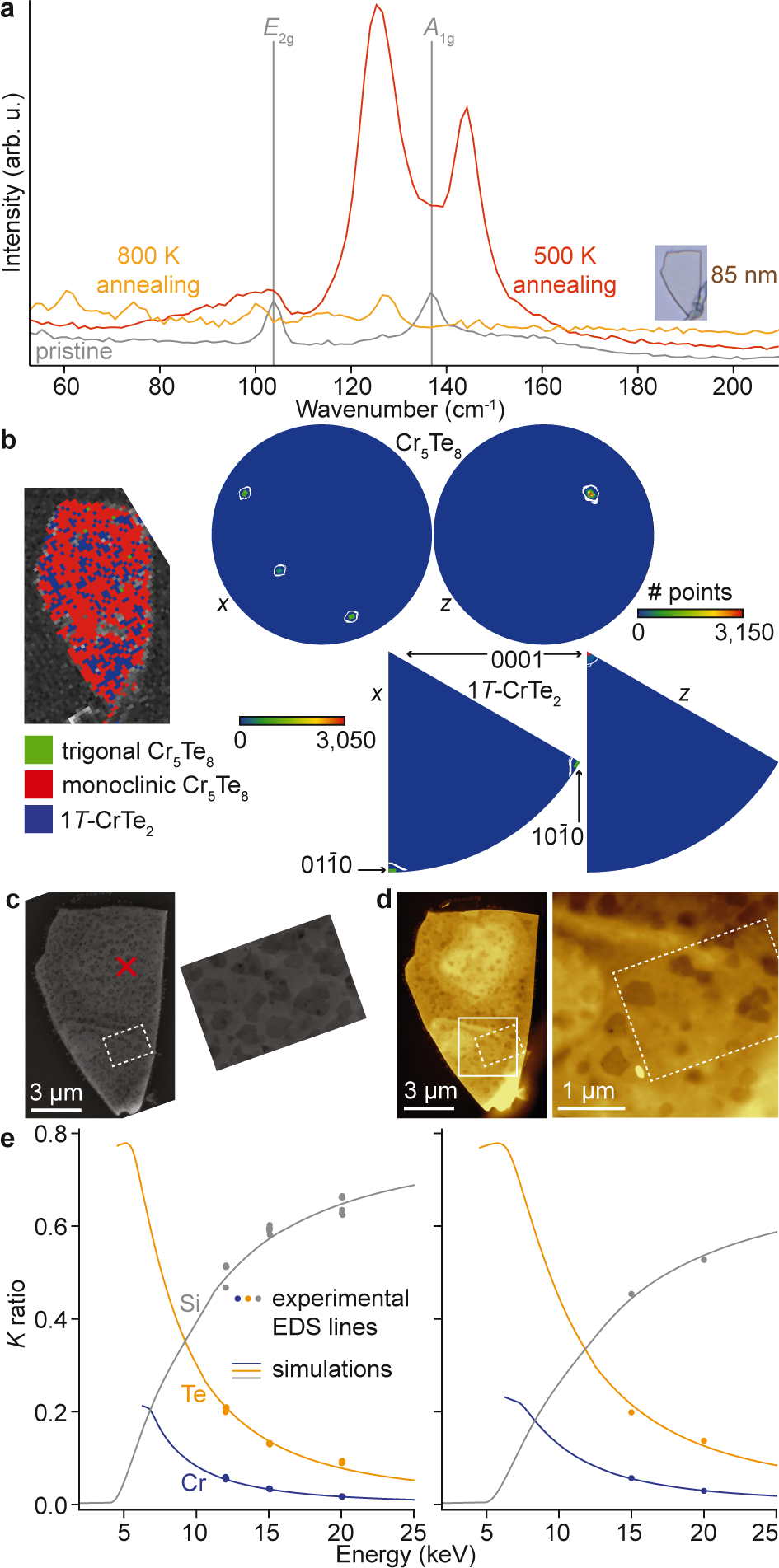}
 \caption{\label{fig2} \textbf{Influence of temperature on structure and composition.} (a) Raman spectra of pristine 1$T$-CrTe$_2$ flakes (85~nm-thickness, same as in \autoref{fig1}) annealed under Ar atmosphere at 500~K and 800~K. (b) Corresponding inverse pole figure extracted from EBSD for 1$T$-CrTe$_2$ and Cr$_5$Te$_8$ within two axes ($x,z$) and map, obtained from SEM-EBSD, of the structure assignment (overlaid on the point-by-point band contrast image), after 800~K annealing. The arrows highlight the projected reflections accumulating most fractions (in number of points) of the SEM images. (c,d) SEM images, acquired with secondary electrons (with a zoom-in within the dotted frame) and corresponding AFM topography images, obtained after annealing to 800~K. (e) EDS analysis of the $K$-ratios for the Cr, Te, Si emission lines (coloured dots) at specific locations marked with crosses on (b), at different electron energies, together with simulated data, after 500~K (left) and 800~K (right) annealing.}
 \vspace{-20pt}
 \end{center}
\end{figure}

The Raman spectrum drastically changes when the pristine material is annealed (see Supporting Information, note in Section~S2). Annealing to 500~K under Ar atmosphere leads to the appearance of two strong peaks, centered at 125.6~cm$^{-1}$ and 144.3~cm$^{-1}$ (\autoref{fig2}a, Supporting Information Figure~S2). The initial $E_{2\text{g}}$ peak is also observed together with a broad feature spanning about 20~cm$^{-1}$ on its low wavenumber side. Besides, to account for the observed line-shapes of the spectra, especially for the relatively strong signal at 136.5~cm$^{-1}$, it seems reasonable that the $A_{1\text{g}}$ peak is also present. This indicates a coexistence of at least two phases in the material, presumably with different Cr:Te composition (see Supporting Information, note in Section~S3). Annealing at 800~K, still under Ar atmosphere, further changes the Raman spectrum. The initial $E_{2\text{g}}$ and $A_{1\text{g}}$ peaks have now disappeared, and peaks centered at 61.4~cm$^{-1}$, 74.9~cm$^{-1}$, 100.5~cm$^{-1}$, 114.0~cm$^{-1}$, 127.4~cm$^{-1}$, and 145.0~cm$^{-1}$ are observed (\autoref{fig2}a, Supporting Information Figure~S2). Some of these new peaks (114.0~cm$^{-1}$, 127.4~cm$^{-1}$, 145.0~cm$^{-1}$) have already been reported for a Cr-Te compound whose composition was estimated, based on a basic analysis of the relative intensity of peaks appearing in EDS, to 5:8 \cite{Fu}. Our Raman spectroscopy data hence shows that annealing pogressively transforms 1$T$-CrTe$_2$ into a compound that could be Cr$_5$Te$_8$ (see sketches of the structure in \autoref{fig0}).

To interpret these observations we apply the careful micro-analysis procedure we used to characterise the structure and composition of pristine 1$T$-CrTe$_2$. After 500~K annealing, the local phase assignment we can perform with SEM-EBSD is not, globally, as straightforward as it was for the pristine flakes. An example is shown in Figure~S3 of the Supporting Information for a 150~nm-thick flake: we resolve two contributions in the inverse pole figures within a single flake. This reflects in the spatially-resolved phase assignment, which still identifies 1$T$-CrTe$_2$ over a large fraction of the flakes, as well as occasionally a Cr$_5$Te$_8$ phase with a fraction of the order of 1\%, and other regions (few 1\% to several 10\%, cf. Figure~S3 of the Supporting Information) with unknown structure (Supporting Information, Section~S7). After 800~K annealing, the situation is clearer: only few 1\% of 1$T$-CrTe$_2$ are left, the rest being transformed into Cr$_5$Te$_8$ (see sketch of the structure in \autoref{fig0}) crystallized in the monoclinic system ($>$ 50\% of the material), to a lesser extent (few 1\%) into trigonal Cr$_5$Te$_8$, and into ($\sim$ 10\%) other (unknown) phases (\autoref{fig2}b). The inverse pole figures for the 1$T$-CrTe$_2$ and monoclinic Cr$_5$Te$_8$ phases reveal that the highest symmetry Cr planes are parallel to the surface (\autoref{fig2}b).

The inhomogeneity is actually directly apparent, and better resolved spatially, in SEM images acquired with secondary electrons (\autoref{fig2}c). We observe lower intensity features with well-defined edges forming 120$^\circ$ angles, all with the same orientation, embedded within a matrix of higher-intensity-material. In the Cr$_5$Te$_8$ structure, the presence of Cr atoms forming bridges between the CrTe$_2$ planes corresponds to a higher atomic density than in 1$T$-CrTe$_2$; hence it should have a higher secondary electron yield and should appear brighter than 1$T$-CrTe$_2$ in the images. We conclude that the dark lower-intensity features we observe are 1$T$-CrTe$_2$ crystallites that have not yet been converted into Cr$_5$Te$_8$. Comparison between SEM images and AFM topographs (\autoref{fig2}d) suggests that these crystallites do not all emerge at the surface, but are also buried below the surface of the flakes.

In addition, the EDS analysis reveals a composition change with the annealing temperature. The change is marginal after 500~K annealing, by typically percents (\autoref{fig2}e). After 800~K annealing, changes are drastic (\autoref{fig2}e). Correlated with the spatially varying secondary electron yield in SEM imaging, the Te/Cr atomic fraction varies spatially, in the form of a bimodal distribution centered at 1.98$\pm$0.02 and 1.72$\pm$0.06.

The former composition is indiscernible from 1:2, which corresponds to 1$T$-CrTe$_2$, but what about the latter composition? A 1.72$\pm$0.06 ratio can point to either a 1$T$-CrTe$_2$ material enriched with Cr atoms, or to 1$T$-CrTe$_2$ depleted with Te, \textit{i.e.} with Te vacancies. Such a depletion would however mean the material assumes essentially the same lattice parameter as 1$T$-CrTe$_2$, in contradiction with our EBSD data (\autoref{fig2}b). The composition derived from EDS hence corresponds to a Cr:Te composition of about 5:8 (actually 5:8.6), with the excess Cr relative to 1$T$-CrTe$_2$ presumably present in the form of covalent Cr bridges connecting the individual 1$T$-CrTe$_2$ layers. Where does this excess Cr come from? No external supply of Cr was provided during the annealing treatments, hence this excess must be either already present or generated locally within the samples. Since X-ray diffraction did not detect any pure-Cr inclusions or other Cr-Te phases in the pristine (macroscopic) grains, a reasonable scenario is Cr diffusion across the sample, leaving Cr-depleted regions likely in the form of Te clusters (we do not detect Cr-poor Cr-Te compound in EBSD).

\subsection*{Room temperature in-plane ferromagnetism in pristine 1$T$-CrTe$_2$ micro-flakes} 
\begin{figure*}[!hbt]
 \begin{center}
 \includegraphics[width=16cm]{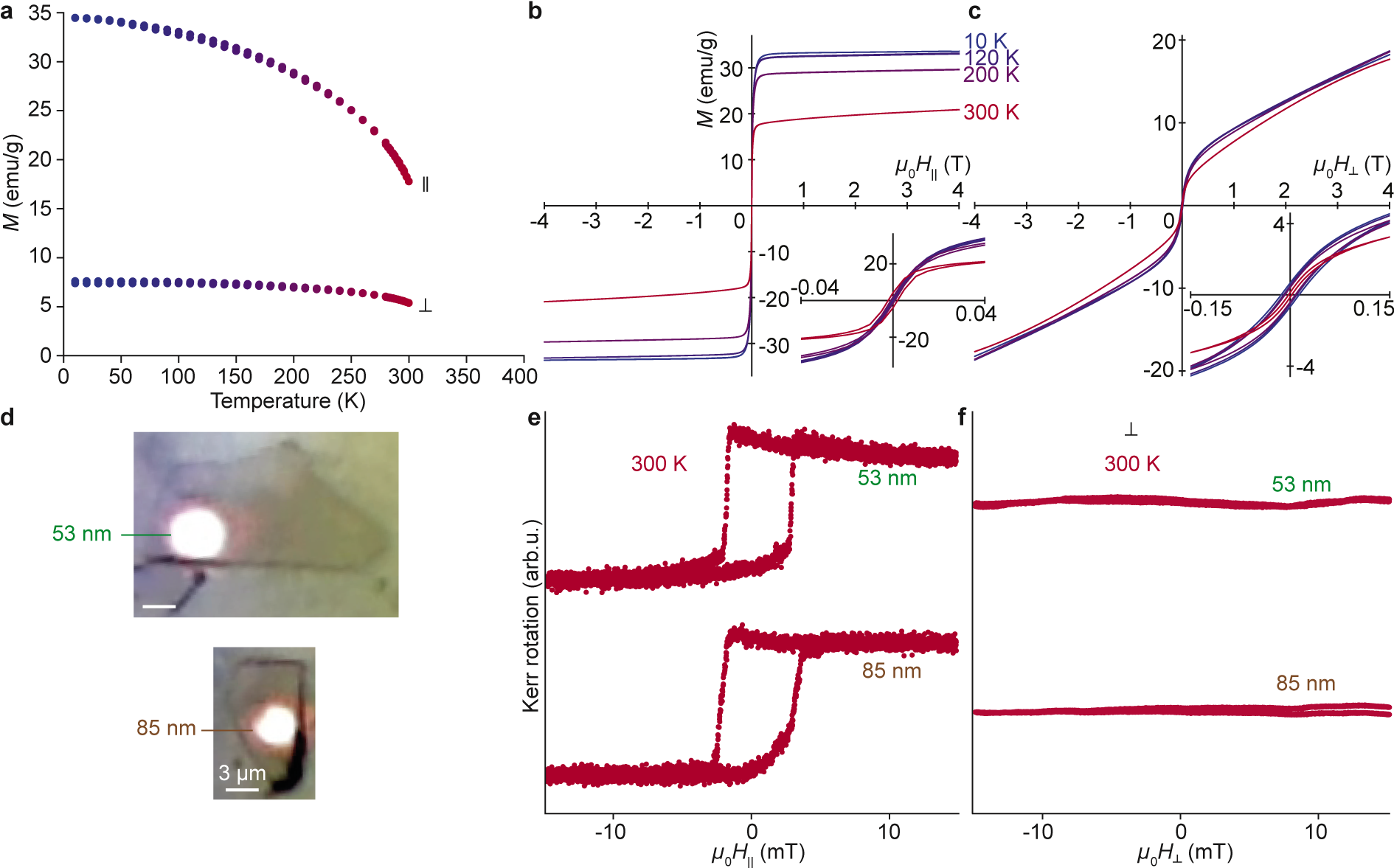}
 \caption{\label{fig3} \textbf{Macroscopic and microscopic magnetic properties of pristine 1$T$-CrTe$_2$.} (a) Magnetisation \textit{versus} temperature, under an applied in-plane ($\parallel$) and out-of-plane ($\perp$) field $\mu_0 H$ = 0.5~T, of a millimeter-sized grain, as measured with a SQUID. (b,c) Corresponding $M(H_{\parallel,\perp})$ in-plane (b) and out-of-plane (c) loops measured at various temperatures; insets are zooms close to the origin. (d) Optical micrographs of the 53~nm- and 85~nm-thick flakes also shown in Figures~\ref{fig1},\ref{fig2} revealing the location of the laser spot used to measure the data in (e,f). (e,f) Corresponding $M(H_{\parallel,\perp})$ local loops measured at room temperature with focused Kerr magnetometry, with an in-plane (e) and out-of-plane (f) magnetic field.}
 \end{center}
\end{figure*}

First, we summarise the temperature dependence of the magnetic properties of millimeter-sized 1$T$-CrTe$_2$ grains, prior to any thermal annealing. Consistent with Ref.~\citenum{Freitas}, magnetization decreases as temperature increases, and has lost $\sim$45\% of its 4~K value close to room temperature, signalling a ferromagnetic-to-paramagnetic transition with $T_\mathrm{Curie}$ = 320~K (\autoref{fig3}a). As expected, the latter does not depend on the magnitude of the external magnetic field or on its direction. However, the maximum magnetization values are substantially larger in the in-plane direction.

The $M(H_{\parallel,\perp})$ magnetization \textit{versus} magnetic field loops with in-plane ($\parallel$) and out-of-plane ($\perp$) applied field have distinct shapes, pointing to an in-plane easy axis for magnetization, regardless of the temperature (\autoref{fig3}b,c). Small loop openings (hence small coercivities) are observed in both directions, but their variations with temperature ($T$) differ: although it increases with decreasing $T$ for $M(H_{\perp})$, it decreases with decreasing $T$ for $M(H_{\parallel})$. Altogether, our observations are very different from those recently reported for what was claimed to be CrTe$_2$ in the form of $\sim$25~nm-thick layers with a $T_\mathrm{Curie}$ = 205~K \cite{Sun_b} and $\sim$25~nm-thick layers with a $T_\mathrm{Curie}$ = 160-190 K~for \cite{Meng}. In Ref.~\citenum{Sun_b} an out-of-plane easy axis was observed, while in Ref.~\citenum{Meng} an in-plane easy was found (beyond the few 1~nm-thickness regime). We argue that the origin of these different magnetic properties are related to different Cr:Te compositions, consistent with the Raman spectroscopy data, beyond the 1:2 ratio that we quantified in the present work with the EBSD + EDS analysis.

Micro-flakes (analyzed with focused Kerr magnetometry at room temperature, \autoref{fig3}d-f) exhibit $M(H)$ loops that look qualitatively different from those of the macroscopic samples (probed with SQUID). The $M(H_\parallel)$ loops reveal a square hysteresis (\autoref{fig3}e), while no loop opening is observed when the field is applied in the perpendicular direction ($M(H_\perp)$, \autoref{fig3}f). This is, once more, typical of an in-plane easy axis for the magnetization (see Supporting Information, Section~S8 and Figure~S4), which is much favoured by the large aspect ratio of the flakes as it minimizes the dipolar term in the total magnetic anisotropy energy.

\subsection*{Low Curie temperatures in thermally-transformed 1$T$-CrTe$_2$} 
\begin{figure}[!hbt]
 \begin{center}
 \vspace{-20pt}
 \includegraphics[width=14.57042cm]{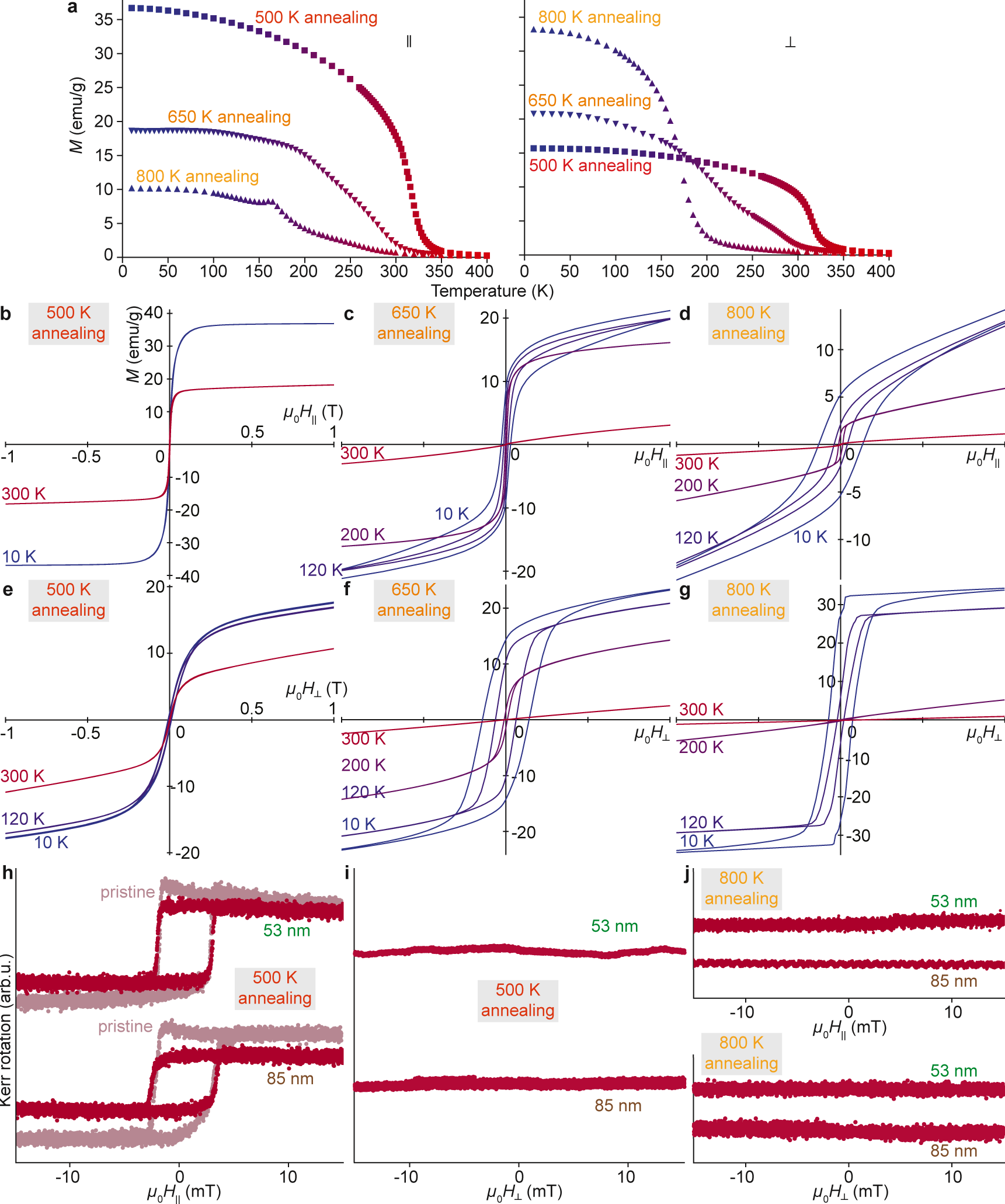}
 \caption{\label{fig4} \textbf{Change of magnetic properties upon thermal annealing.} (a) Magnetisation \textit{versus} temperature under a $\mu_0 H_{\parallel,\perp}$ = 0.5~T field applied in-plane and out-of-plane, to a millimeter-sized grain annealed to 500~K, 650~K, and 800~K under Ar, as measured with a SQUID. (b-g) Corresponding in-plane (b-d) and out-of-plane (e-g) $M(H_{\parallel,\perp})$ loops measured at various temperatures. (h,i) $M(H_{\parallel,\perp})$ loops measured for the 53~nm- and 85~nm-thick flakes (also studied in Figures~\ref{fig1},\ref{fig2},\ref{fig3}) with focused Kerr magnetometry, with a magnetic field applied in-plane (h and top-two curves in j) and out-of-plane (i and bottom-two curves in j). The steps observed in the 85~nm-thick sample in (i) are artefacts, stemming from a non perfectly perpendicular orientation of the applied magnetic field.}
 \vspace{-20pt}
 \end{center}
\end{figure}

Millimeter-sized grains were subsequently annealed to 500~K, 650~K, and 800~K under Ar atmosphere. The $M(T)$ curves (\autoref{fig4}a) reveal a progressive decrease of $T_\mathrm{Curie}$, well below 300~K. While a well-defined $T_\mathrm{Curie}$ is still observed after 500~K annealing (see $M(T)$ derivative in Figure~S5 of the Supporting Information), it is noteworthy that the perpendicular component of magnetization has doubled compared to pristine samples. After annealing to 650~K, the perpendicular component exceeds the parallel component; additionally multiple steps are observed in the $M(T)$ curves, at $\sim$130~K, $\sim$210-220~K and $\sim$280~K suggesting a coexistence of phases with distinct $T_\mathrm{Curie}$ (Supporting Information, Figure~S5). Finally, after 800~K annealing, the perpendicular component largely dominates and exhibits a monotonous decay with $T$ pointing to $T_\mathrm{Curie}\sim$175~K (Supporting Information, Figure~S5). Instead, a more complex behaviour is observed for the parallel component, which exhibits a bump at 165~K and a step at $\sim$250~K.

The unambiguous changes of the magnetic easy axis (from in-plane to out-of-plane) and of the ordering temperature from above to below room temperature are also evident in the $M(H_{\parallel,\perp})$ loops measured at various temperatures (\autoref{fig4}b-g; see also Figure~S6 in Supporting Information for a broader range of field values).

To interpret these observations we now refer to our characterizations of the structure and composition (\autoref{fig2}) and to data available from previous literature. Our SEM EBSD/EDS analysis suggests that a negligible amount of the pristine 1$T$-CrTe$_2$ material has transformed after 500~K annealing, turning into other phases such as CrTe and Cr$_5$Te$_8$. This is consistent with our Raman spectroscopy data, which appears to be especially sensitive to these other phases, and with the $M(T)$ and $M(H)$ curves, which are only slightly altered by this moderate annealing.

Changes are stronger after 650~K annealing, with steps in the $M(T)$ curves that are consistent with the presence of CrTe crystallites (whose $T_\mathrm{Curie}$ varies with their thickness \cite{Wang_b}) or Cr$_5$Te$_8$ \cite{Wang_d}. A large fraction of the material then consists of phases other than 1$T$-CrTe$_2$.

Finally, after 800~K annealing, our previous analyses indicate that the material is to a large extent converted to Cr$_5$Te$_8$, still with few 100~nm-wide 1$T$-CrTe$_2$ clusters. Obviously, these clusters do not significantly contribute to the magnetic response of the material. They represent a small fraction of the material, typically few 1\% (see discussions above), which may still be sufficient to be detected with SQUID magnetometry. A possible reason is that their low thickness (\autoref{fig2}c,d shows they are definitely thinner than the flakes) corresponds to a reduced $T_\mathrm{Curie}$ \cite{Zhang_b}; they hence may contribute to the $>$ 200~K tail of the $M(T)$ curve measured with an in-plane field (\autoref{fig4}a, left). Anyway, the observed ordering temperature of the 800~K-annealed material, around 175~K, is consistent with our above assignment (\autoref{fig2}) of a Cr$_5$Te$_8$ structure, for which similar $T_\mathrm{Curie}$ values were recently reported for a certain range of material thickness \cite{Tang}. Our samples may indeed consist of Cr$_5$Te$_8$ with thickness of the order of 10~nm, separated by other kinds of more-or-less well-defined Cr-Te phases. In contrast, bulk Cr$_5$Te$_8$ exhibits a higher ordering temperature, $T_\mathrm{Curie}$ $\sim$ 230~K \cite{Wang_d}. Besides, similar to the observation made for the bulk compound, Cr$_5$Te$_8$ in our study exhibits a bump in the $M(T)$ for the parallel component, although at a lower temperature (165~K). The origin of this bump is not yet understood, possibly related to the presence of non-trivial spin textures \cite{Wang_d}.

Finally, we discuss the room temperature magnetism of the flakes studied before and after annealing in Figures~\ref{fig1},\ref{fig2},\ref{fig3}. We find square hysteresis loops in the parallel direction after 500~K annealing, and no loop opening in the perpendicular direction. This is qualitatively similar to the case of the pristine flakes (\autoref{fig4}h,i), and we observe a significant reduction of the intensity of the Kerr rotation after 500~K annealing, confirming once more that part of the pristine 1$T$-CrTe$_2$ has been converted into another phase. Strikingly, after annealing to 800~K, no ferromagnetic signature can be detected in either direction (\autoref{fig4}j), in perfect agreement with SQUID measurements revealing magnetic ordering only at low temperatures.

\section*{Summary and conclusion}
Using detailled structural and chemical analysis of micro-flakes obtained by mechanical exfoliation from a bulk material synthesized at high temperature, we have established that these micro-flakes are made of the room temperature 1$T$-CrTe$_2$ ferromagnet. We have presented their polarization-dependent Raman spectroscopy signature, and showed that this is a very powerfull tool to discriminate the different Cr-Te compounds. We next have presented their magnetic properties, namely an in-plane easy axis, a Curie temperature above 300~K (at least for several 10~nm-thick samples), and coercivities of the order of few 1~mT. We have found that the bulk material, and flakes derived from it, are stable over weeks in air and at room temperature, and we have detected neither signs of oxidation or other chemical transformations, nor changes in the crystal structure, with Raman spectroscopy, electron diffraction, energy dispersive spectroscopy, optical Kerr and SQUID magnetometries.

We have then annealed the bulk 1$T$-CrTe$_2$ material and its micro-flakes at temperatures of 500~K, 650~K, and 800~K. We have found that the structure and composition of the starting material changes upon these treatments, rather marginally at 500~K but prominently starting from 650~K. The material converts, to a large extent, to monoclinic Cr$_5$Te$_8$, a phase where the CrTe$_2$ planes are bridged by Cr atoms. We have correlated the changes of structure and composition to profound changes of the magnetic properties, which are, after 800~K, characterized by an out-of-plane anisotropy and a Curie temperature below 200~K. We suggest that the origin of the strong change in the magnetic properties is rooted in the role of covalent Cr-Te interlayer bonds, which may introduce super-exchange kinds of magnetic interactions between the transition metal ions belonging to different CrTe$_2$ layers in Cr$_5$Te$_8$.

These findings provide a natural guide to revisit the recent literature (and to analyse commercially available samples, see Supporting Information Figure~S7), which witnessed a number of conflicting reports regarding the properties of what was claimed to be 1$T$-CrTe$_2$. Besides shedding new light on the compound properties, our work highlights new opportunities for thermally-induced phase engineering. Given the richness of magnetic properties of Cr-Te compounds, ultra-thin materials with a variety of remarkable properties are within reach, \textit{e.g.} hosting room-temperature ferromagnetism \cite{Purbawati}, topologically non-trivial magnetic textures \cite{Li,Saha} or a very strong anomalous Hall effect \cite{Tang}, and could be combined within in-plane phase-modulated materials \cite{Li_b,Yao} by exploiting local heating effects.

\section*{Materials and Methods}

1$T$-CrTe$_2$ samples were obtained from KCrTe$_2$. KCrTe$_2$ was prepared from a molar mixture of Cr, K and Te, introduced under argon atmosphere (glove box) within an evacuated quartz tube. The quartz tube was then heated to the melting point of K and Te, and held at 1170~K during eight days. After cool down, the tubes were opened in a glove box to prevent oxidation. Potassium de-intercalation was subsequently achieved by reaction of KCrTe$_2$ with a solution of iodine in acetonitrile. The resulting suspensions were stirred for about 1~h in an excess of iodine. Finally, washing with acetonitrile to remove iodide, filtering and drying under vacuum produce gray shinny platelets of lateral extension of few millimeters and few 100~$\mu$m thickness.

These platelets were exfoliated onto thin Pt films (1~nm tickness) deposited onto Ta thin films (10~nm thickness) sputtered on Si wafers. These substrates conveniently evacuate the electron current induced during measurements performed under the electron beam of the SEM, and mitigate charging effects that otherwise make measurements very cumbersome. The samples were analyzed prior to any thermal treatment (pristine) and after successive thermal annealings. The time spent in atmospheric conditions was minimized to typically few hours.

Focused Kerr magnetometry was performed using a home-made setup. This setup includes a He-Ne laser (632~nm) with 0.3~mW power and a 100$\times$ objective focusing the laser beam with $s$ (linear) polarization to a 1~$\mu$m diameter spot on the sample surface. The reflected beam goes through a Wollaston prism beam splitter producing two beams, with orthogonal polarization, and whose intensity is measured with two identical photodiodes. An oscilloscope is used to acquire the sum of the two corresponding signals during sweeps (1-2~Hz frequency, 100 sweeps typically used to improve the signal-to-noise ratio) of an external applied magnetic field. The field was applied using two kinds of electromagnets, a small horseshoe one and a linear solenoid, producing in-plane ($\parallel$) and out-of-plane ($\perp$) magnetic fields, respectively. The perpendicular loops were measured in polar MOKE configuration, with the incoming and reflected laser beam perpendicular to the sample surface. The in-plane loops were measured by moving the laser beam out of the center of the objective, leading to an angle of about 20$^\circ$ between the incoming laser beam and the normal to the sample surface. This is thus not an optimal LMOKE configuration. In order to avoid polar and longitudinal Kerr signals in this geometry it was important to accurately align the applied in-plane magnetic field. With only an in-plane field, the component of the magnetization perpendicular to the surface should not change upon sweeping the field and no varying polar Kerr signal should be observed. For the data shown in \autoref{fig3} and \autoref{fig4}, a linear contribution \textit{versus} applied magnetic field, corresponding to the Faraday rotation through the optics, has been subtracted.

Raman spectra were acquired using a Witec Alpha 500 Raman microscope with a 532~nm laser (power set to 0.4~mW to ensure that samples are not damaged) focused with a 50$\times$ objective (Mitutoyo, NA=0.75) to a 1~$\mu$m-wide spot. To access to low-wavenumber ranges such as needed to characterise $1T$-CrTe$_2$, the optical detection line was equipped with a Rayshield coupler. The spectrometer was mounted with a 1800~lines/mm grating, with which a spectral resolution $\leq$ 0.1 cm$^{-1}$ was possible. Acquisition of the spectra was done with typically 60~s integration time and typically three accumulations. Measurements were performed at room temperature for samples sealed within Ar-filled quartz cells (\autoref{fig1}f,g) and samples having experienced various thermal treatments, outside of the Ar-filled quartz cells.

SEM imaging, combined with EBSD and EDS micro-analyses, were performed over about 10 full exfoliated flakes with a field-emission-gun FE-SEM Zeiss Ultra Plus instrument (see more details in Supporting Information Section~S5). Energy-dispersive X-ray data were acquired on several locations of each flake, with a Bruker silicon drift detector (SDD) using electron energies of 12~keV, 15~keV and 20~keV. Secondary electron imaging was performed with an Everhart \& Thornley detector at 15~keV. Quantitative micro-analysis was achieved by comparing the X-ray emission line intensities to those measured with standard materials, namely pure Cr, Te, Si, Pt Ta, and for K using the KTiOPO$_4$ compound. The data were analysed using the StrataGEM software \cite{Pouchou}, taking into account the presence of the substrate and deriving the composition and flake thickness by comparison of the $K$-ratios as function of electron beam energy to simulated values. To obtain satisfactory fits of the simulated values to the experimental $K$-ratios, no global scale factor was needed and only the thickness and composition of the flakes and substrate were free fit parameters. The flake thicknesses evaluated from this analysis were found to be very consistent with the AFM estimates. For EBSD data-processing and analysis, Oxford softwares (Tango \& Mambo) were used.

Atomic force microscopy was done at room temperature in ambiant conditions, using a Veeco Dimension 3100 apparatus operated in tapping mode with OMCL-AC160TS probes.

\section{Author Information}
*E-mail: johann.coraux@neel.cnr.fr

\section{Associated Content}
The authors declare no competing financial interest.

\begin{acknowledgement}

We thank Simon Le Denmat for assistance in the AFM measurements. We warmly thank Michel Hehn for providing the Pt/Ta/Si substrates. This work was supported by the Agence Nationale de la Recherche through projects No. ANR-17-CE24-0007-03 `Bio-Ice' and ANR-19-CE24-0021 `ANETHUM', and under the program ESR/EquipEx+ (grant number ANR-21-ESRE-0025). 

\end{acknowledgement}

\begin{suppinfo}

Supporting information comprises a Raman spectroscopy analysis as function of the annealing temperature for one of the two flakes analyzed in \autoref{fig1}, an AFM and SEM-EBSD study of an additional flake after 500~K annealing, an analysis of the vibrational, composition and magnetic properties of a commercially-available material, and a survey of the possible Cr-Te compounds and their structure/composition.

\end{suppinfo}



\providecommand{\latin}[1]{#1}
\makeatletter
\providecommand{\doi}
  {\begingroup\let\do\@makeother\dospecials
  \catcode`\{=1 \catcode`\}=2 \doi@aux}
\providecommand{\doi@aux}[1]{\endgroup\texttt{#1}}
\makeatother
\providecommand*\mcitethebibliography{\thebibliography}
\csname @ifundefined\endcsname{endmcitethebibliography}
  {\let\endmcitethebibliography\endthebibliography}{}

\end{document}